\documentclass{ws-procs9x6}

\def\al{\alpha}

\def\ga{\gamma}

\def\ka{\kappa}
\def\la{\lambda}

\def\si{\sigma}

\def\ta{\tau}

\def\ps{\psi}
\def\om{\omega}
\def\Ga{\Gamma}

\def\mn{{\mu\nu}}

\def\half{{\textstyle{1\over 2}}}

\def\frac#1#2{{\textstyle{{#1}\over {#2}}}}

\def\lsim{\mathrel{\rlap{\lower4pt\hbox{\hskip1pt$\sim$}}
    \raise1pt\hbox{$<$}}}
\def\gsim{\mathrel{\rlap{\lower4pt\hbox{\hskip1pt$\sim$}}
    \raise1pt\hbox{$>$}}}
\def\sqr#1#2{{\vcenter{\vbox{\hrule height.#2pt
         \hbox{\vrule width.#2pt height#1pt \kern#1pt
         \vrule width.#2pt}
         \hrule height.#2pt}}}}

\def\etal{{\it et al.}}
\def\ibid{{\it ibid}}

\def\pt#1{\phantom{#1}}

\def\ol#1{\overline{#1}}

\def\nsc#1#2#3{\om_{#1}^{{\pt{#1}}#2#3}}

\def\ivb#1#2{e^{#1}_{{\pt{#1}}#2}}
\def\uvb#1#2{e^{#1#2}}

\def\ab{\overline{a}}

\def\twiddle{\lower4pt\hbox{\hskip-0pt{$\widetilde{}$}}}
\def\m@th{\mathsurround=0pt}
\def\cmapstochar{\mathrel{\rlap{
  \lower0.1pt\hbox{\hskip-1.75pt{$\mapstochar$}}}
  \raise0pt\hbox{\hskip2.5pt{$\twiddle$}}}}
\def\notsimfill{$\m@th\cmapstochar$}
\def\scroodle#1{\vbox{\ialign{##\crcr\notsimfill\crcr
  \noalign{\kern-4pt\nointerlineskip}
   $\hfil\displaystyle{#1}\hfil$\crcr}}}

\def\cmapstocharbig{\mathrel{\rlap{
  \lower0.1pt\hbox{\hskip0.25pt{$\mapstochar$}}}
  \raise0pt\hbox{\hskip4.5pt{$\twiddle$}}}}
\def\notsimfillbig{$\m@th\cmapstocharbig$}
\def\scroodlebig#1{\vbox{\ialign{##\crcr\notsimfillbig\crcr
  \noalign{\kern-4pt\nointerlineskip}
   $\hfil\displaystyle{#1}\hfil$\crcr}}}

\def\X{t_{\la \mn \ldots}}

\def\Xb{\overline{t}_{\la \mn \ldots}}

\def\Xtw{\scroodle{t}_{\la \mn \ldots}}

\def\af{(a_{\rm{eff}})}

\def\afb{(\ab_{\rm{eff}})}

\def\atw{\scroodle{a}{}}

\def\aftw{(\scroodle{a}_{\rm{eff}})}

\def\hnr{H_{\rm NR}}

\def\b2{b^\al b_\al}

\def\lrpartial{\raise 1pt\hbox{$\stackrel\leftrightarrow\partial$}}

\def\lrDmu{\stackrel{\leftrightarrow}{D_\mu}}

\def\a{$a_\mu$ }
\def\b{$b_\mu$ }

\def\e{$e_\mu$ }

\newcommand{\beq}{\begin{equation}}
\newcommand{\eeq}{\end{equation}}
\newcommand{\bea}{\begin{eqnarray}}
\newcommand{\eea}{\end{eqnarray}}
\newcommand{\bit}{\begin{itemize}}
\newcommand{\eit}{\end{itemize}}

\begin{document}

\title{Probing Lorentz Symmetry with Gravitationally Coupled Matter}

\author{JAY D.\ TASSON}

\address{Physics Department, Indiana University\\ 
Bloomington, IN 47405, USA\\ 
E-mail: jtasson@indiana.edu}

\maketitle

\abstracts{
Methods for obtaining
additional sensitivities to Lorentz violation
in the fermion sector of the Standard-Model Extension
using gravitational couplings are discussed.}

\section{Introduction}

The Standard Model of particle physics
together with Einstein's General Relativity
provide a remarkably successful description
of known phenomena.
General Relativity describes gravitation
at the classical level,
while all other interactions
are described down to the quantum level
by the Standard Model.
However,
a single quantum-consistent theory
at the Planck scale
remains elusive.

Ideally,
experimental information
would guide the development
of the underlying theory;
however,
directly probing the Planck scale
is impractical
at present.
A feasible alternative
is to search for suppressed effects
arising from Planck-scale physics
in sensitive experiments
that can be performed at presently accessible energies.
Relativity violations
arising from Lorentz-symmetry violation
in the underlying theory
provide a candidate suppressed effect.\cite{cpt07,sme1}
The Standard-Model Extension (SME)
is an effective field theory
that describes Lorentz violation
at our present energies.\cite{sme,akgrav}

A large number of experimental tests
of Lorentz symmetry
have been performed
in the context of the minimal SME.
Those test include,
in the Minkowski-spacetime limit,
experiments with
electrons,\cite{eexpt}
protons and neutrons,\cite{pnexpt}
photons,\cite{photonexpt}
mesons,\cite{hadronexpt}
muons,\cite{muexpt}
neutrinos,\cite{nuexpt}
and the Higgs.\cite{higgs}
The pure-gravity sector
has is also being investigated
in the post-Newtonian limit.\cite{qagrav,grexpt}
Although no compelling
experimental evidence
for Lorentz violation
has been found to date,
much remains unexplored.
For example,
only about half
of the coefficients for Lorentz violation
involving light and ordinary matter
(protons, neutrons, and electrons)
have been investigated experimentally,
and other sectors
remain nearly unexplored.

In the remainder of this proceedings,
a theoretical basis
for extending (SME) studies with ordinary matter
into the post-Newtonian regime,
developed with Alan Kosteleck\'y,
will be discussed.\cite{lvgap}
The goal of this work
is to obtain new sensitivities to Lorentz violation
in the fermion sector
using couplings to gravity.
These couplings
introduce new operator structures
that provide sensitivities
to coefficients for Lorentz violation
that are unobservable
in Minkowski spacetime.

\section{Relativistic Theory}

Gravitational effects
are incorporated into the SME action
in Ref.\ \refcite{akgrav}.
The general geometric framework assumed
is Riemann-Cartan spacetime, 
which allows for a nonzero torsion tensor $T^\la_{\pt{\la} \mu \nu}$
as well as
the Riemann curvature tensor $R^\ka_{\pt{\ka} \la \mu \nu}$.
Due to the need to incorporate spinor fields,
the vierbein formalism is adopted.
The vierbein also allows
one to easily distinguish general coordinate
and local Lorentz transformations,
a feature convenient in studying Lorentz violation.\cite{akgrav}
The spin connection $\nsc \mu a b$
along with the vierbein $\ivb \mu a$
are taken as the fundamental
gravitational objects,
while the basic non-gravitational fields
are the photon $A_\mu$
and the Dirac fermion $\ps$.

The minimal-SME action
can be expanded in the following way:
\beq
S = S_G + S_\ps + S^\prime.
\label{SMEaction}
\eeq 
Here, 
first term, $S_G$
is the action 
of the pure-gravity sector,
which contains the dynamics
of the gravitational field
and can also contain
coefficients for Lorentz violation
in that sector.\cite{akgrav,qagrav}
The Einstein-Hilbert action
of General Relativity
is recovered in the limit of
zero torsion
and Lorentz invariance.
The action for the fermion sector
is provided by the term $S_\ps$
in Eq.\ (\ref{SMEaction}).
\beq
S_\ps
= 
\int d^4 x (\half i e \ivb \mu a \ol \ps \Ga^a \lrDmu \ps 
- e \ol \ps M \ps). 
\label{qedxps}
\eeq
Here,
$\Ga^a$ and $M$
take the form shown in the following definitions.
\bea
\Ga^a
&\equiv & 
\ga^a - c_{\mu\nu} \uvb \nu a \ivb \mu b \ga^b
- d_{\mu\nu} \uvb \nu a \ivb \mu b \ga_5 \ga^b
\nonumber\\
&&
- e_\mu \uvb \mu a 
- i f_\mu \uvb \mu a \ga_5 
- \half g_{\la\mu\nu} \uvb \nu a \ivb \la b \ivb \mu c \si^{bc}.
\\
M
& \equiv &
m + a_\mu \ivb \mu a \ga^a 
+ b_\mu \ivb \mu a \ga_5 \ga^a 
+ \half H_{\mu\nu} \ivb \mu a \ivb \nu b \si^{ab} .
\label{mdef}
\eea
The symbols
$a_\mu$, $b_\mu$, $c_\mn$, $d_\mn$, $e_\mu$, $f_\mu$, $g_{\mn \la}$, $H_\mn$
are the coefficient fields for Lorentz violation
of the minimal fermion sector.
In general,
they vary with position
and
differ for each species of particle.
For additional discussion
of the fermion-sector action,
see Ref.\ \refcite{akgrav}.

The final portion, 
$S^\prime$, 
of the action (\ref{SMEaction})
contains the dynamics associated 
with the coefficient fields for Lorentz violation
and is responsible for spontaneous breaking of Lorentz symmetry.
Through symmetry breaking,
the coefficient fields for Lorentz violation
are expected to
acquire vacuum values.
Thus,
it is possible to write
\beq
\X = \Xb + \Xtw,
\eeq
where $\X$
represents an arbitrary coefficient field for Lorentz violation,
$\Xb$ is the corresponding vacuum value,
and $\Xtw$
is the fluctuations
about that vacuum value.
Note that a subset of these fluctuations
are the massless Nambu-Goldstone modes
associated with Lorentz-symmetry breaking.\cite{ngmodes}
It is possible to develop
the necessary tools to analyze fermion experiments
in the presence of gravity and Lorentz violation
without specifying $S^\prime$.\cite{lvgap}
Those results are summarized here.

The relativistic quantum-mechanical hamiltonian
obtained from action (\ref{qedxps})
provides a first step
toward the goal
of obtaining experimental access to Lorentz violation.
The hamiltonian my be obtained pertrubatively
since gravitational and Lorentz-violating effects
are small in the regimes of interest.
Gravity may be considered perturbatively small
in the laboratory and solar-system tests
that will be of interest,
and Lorentz violation is assumed small
since it has not been observed in nature.
In what follows,
orders in these small quantities
will be denoted $O(m,n)$,
where $m$ and $n$
are the order of the given term
in coefficients for Lorentz violation
(vacuum values)
and the metric fluctuation
respectively.

After incorporating relevant effects to order $m+n=2$
and making a field redefinition required to define the hamiltonian,\cite{lvgap,acham}
a hamiltonian of the form,
\beq
H = H^{(0,0)} + H^{(0,1)} + H^{(1,0)} + H^{(1,1)} + H^{(0,2)},
\eeq
is found.
Here, the Lorentz invariant contributions consist of
the conventional Minkowski-spacetime hamiltonian, $H^{(0,0)}$,
and the first and second order gravitational corrections
denoted $H^{(0,1)}$ and $H^{(0,2)}$ respectively.
The first order correction
to the hamiltonian
due to Lorentz violation,
$H^{(1,0)}$,
is the same as that found 
in the flat-spacetime SME.\cite{acham}
The $O(1,1)$ correction
to the hamiltonian
is the term of interest
since it contains
coefficients for Lorentz violation coupled to gravity.
It can be written as follows:
\beq
H^{(1,1)} = H^{(1,1)}_h + H^{(1,1)}_a + H^{(1,1)}_b + \dots  + H^{(1,1)}_g + H^{(1,1)}_H,
\eeq
where $H^{(1,1)}_h$ is the perturbation to the hamiltonian
from Lorentz-violating corrections
to the metric.
The relevant corrections to the metric
can be obtained though investigations
of the classical limit in Sec.\ \ref{classicaltheory}.
The terms denoted $H^{(1,1)}_t$
are perturbations to the hamiltonian
due to Lorentz-violating effects
on the test particle at $O(1,1)$.
As a sample of $O(1,1)$ effects,
$H^{(1,1)}_a$ takes the following form:
\beq
H^{(1,1)}_a =
\atw_0 
- \ab^j h_{j0}
+ \Big[ \atw_j
- \half \ab_j h_{00}
- \half \ab^k h_{jk} \Big] \ga^0 \ga^j.
\eeq
The explicit form of the relativistic hamiltonian
including all of the coefficients of the minimal fermion sector of the SME,
along with additional details of its derivation,
can be found in Ref.\ \refcite{lvgap}.

\section{Non-relativistic Theory}
\label{nonrel}

The relativistic hamiltonian above
is most useful as a tool
to derive the non-relativistic hamiltonian, 
rather than to analyze 
experiments directly,
because experiments most sensitive
to the position operator
are non-relativistic.
A Foldy-Wouthuysen transformation
can be applied to obtain
the non-relativistic hamiltonian
taking care to maintain the desired order
in the small quantities.

After performing the transformation,
the non-relativistic hamiltonian, $\hnr$,
can be written,
\beq
\hnr = \hnr^{(0,0)} + \hnr^{(0,1)} + \hnr^{(1,0)} + \hnr^{(1,1)}+ \hnr^{(0,2)}.
\label{nrh}
\eeq
Here,
as in the relativistic case,
$\hnr^{(0,0)}$ is the conventional Minkowski-spacetime hamiltonian,
$\hnr^{(0,1)}$ and $\hnr^{(0,2)}$ contain the leading and sub-leading  gravitational corrections,
and $\hnr^{(1,0)}$ are the leading corrections
due to Lorentz violation,
which match Ref.\ \refcite{acham}.
Again, the leading couplings of Lorentz violation
to gravitational effects, $\hnr^{(1,1)}$,
are the contributions of interest.
In a manner analogous to the relativistic case,
this term contains contributions 
from the modified metric fluctuation,
written $H_{{\rm NR},h}^{(1,1)}$,
as well as perturbations 
from each of the coefficients for the test particle, $H_{{\rm NR},t}^{(1,1)}$.

A result of the Foldy-Wouthuysen transformation
is that at leading order
in the coefficients for Lorentz violation
at the non-relativistic level,
coefficients $a_\mu$ and $e_\mu$
always appear in the combination
$\af_\mu \equiv a_\mu - m e_\mu$.
This result is consistent
with those of Ref.\ \refcite{akgrav}
indicating that $e_\mu$
can be redefined into $a_\mu$
at leading order in the coefficients for Lorentz violation
via an appropriate
field redefinition.
As a sample of $O(1,1)$ contributions to $\hnr$,
the correction
to the hamiltonian
due to \a and \e
can be written
\beq
H_{{\rm NR}, a_{\rm eff}}^{(1,1)} = \aftw_0 
+ \afb_k h^{0k} 
- \frac{1}{m} \afb^j h_{jk} p^k
+ \frac{1}{m} \Big[ \aftw_j
- \half \afb_j h_{00} \Big] p^j,
\eeq
to second order in momentum.
Additional details of the derivation of $\hnr$,
along with an explicit form 
for the remaining spin-independent contributions to $\hnr$
can be found in Ref.\ \refcite{lvgap}.

\section{Classical Theory}
\label{classicaltheory}

The classical action can be written
as the sum of partial actions
\beq
S = S_G + S_u + S^\prime,
\eeq
just as in the relativistic case.
Here, 
$S_G$ and $S^\prime$
are as in Eq.\ (\ref{SMEaction}).
The partial action $S_u$ 
is the point-particle limit 
of $S_\ps$ appearing in Eq.\ (\ref{SMEaction}).
The classical theory
is useful for several applications including
obtaining
the equations of motion
for a classical particle,
analyzing
non-relativistic quantum-mechanical experiments
via the path integral approach,
and obtaining the modified metric.

Upon inspection
of the non-relativistic hamiltonian
discussed in Sec.\ \ref{nonrel},
a point-particle action
which corresponds to this hamiltonian
can be found.
The spin-independent contributions 
to this action action
can be written
\beq
S_u = \int d\ta \left(-m  \sqrt{-(g_\mn + 2c_\mn) u^\mu u^\nu}
-(a_{\rm eff})_\mu u^\mu \right),
\label{actionsec}
\eeq
where $u^\mu$ is the four velocity
as usual.
The validity of this action
has been established here
only to within the assumptions made
up to the presentation of the non-relativistic hamiltonian, $\hnr$,
and in performing calculations 
with this action
one should not exceed
the order in small quantities
or the order in momentum
discussed in the last section.
In addition to the match between the classical action above
and the non-relativistic hamiltonian,
the fact that the dispersion relation
generated by the classical action
matches the relativistic theory
confirms the validity of this action.
Note also
that the $c_\mn$ contributions to this action
have been discussed previously
in the context of work done
in the photon sector.\cite{qbem}
After some consideration,\cite{lvgap}
this action can be extended
to address the case
in which particles are bound
within macroscopic matter
as well.

\section{Experimental Tests}

With the theory developed above,
experiments in any regime,
from relativistic quantum mechanics
to classical mechanics,
can be analyzed,
provided the contributions from $S^\prime$
are known.
These contributions
can be established directly
within a model of spontaneous Lorentz violation
or determined for a large class of models
by examining the general form of the contributions
along with the constraints available from conservation laws.

Upon a general analysis,
effects
are found in a number of experiments.
These effects include annual and sidereal variations
in the newtonian gravitational acceleration
as well as variations in the gravitational force
based on the proton, neutron, electron content
of the bodies involved.
These effects
lead to signals
in gravimeter tests
as well as in some experiments
designed to test the weak equivalence principle.
These tests
are described in detail
in Ref.\ \refcite{lvgap}
and will provide the first direct sensitivities
to the $\afb_\mu$ coefficients
for the proton, neutron, and electron.

\end{document}